\documentclass[prd,
eqsecnum,floatfix,letterpaper,superscriptaddress,groupedaddress,nofootinbib]{revtex4}

\usepackage{amsmath}
\usepackage{latexsym}
\usepackage{amssymb}
\usepackage{amsfonts}
\usepackage{mathtools}
\usepackage{bm}
\usepackage{amsthm}
\usepackage{graphicx}
\usepackage{color}
\usepackage{changes}
\usepackage[normalem]{ulem}
\usepackage{natbib}
\def\no{\nonumber}
\def \D{{\mathcal D}}

\def \a {\alpha}
\def \b {\beta}
\def \tpp {t^{\prime \prime}}

\def \be{\begin{equation}}
\def \bea{\begin{eqnarray}}
\def \eea{\end{eqnarray}}
\def \ee{\end{equation}}

\begin{document}
\title{Reply to the Bayle {\it et al.} gr-qc document dated June 7, 2021}
\author{Massimo Tinto}
\email{mtinto@ucsd.edu}
\affiliation{University of California San Diego,
  Center for Astrophysics and Space Sciences,
  9500 Gilman Dr, La Jolla, CA 92093,
  U.S.A.}
\affiliation{Divis\~{a}o de Astrof\'{i}sica, Instituto
  Nacional de Pesquisas Espaciais, S. J. Campos, SP 12227-010, Brazil}
\author{Sanjeev Dhurandhar}
\email{sanjeev@iucaa.in}
\affiliation{Inter University Centre for Astronomy and Astrophysics,
  Ganeshkhind, Pune, 411 007, India}
\author{Prasanna Joshi}
\email{joshi.prasanna@students.iiserpune.ac.in}
\affiliation{IISER, Pashan, Pune, 411 008, India}

\date{\today}

\begin{abstract}
  We address the two issues raised by Bayle, Vallisneri, Babak, and
  Petiteau (in their gr-qc document arxiv.org/abs/2106.03976) about
  our matrix formulation of Time-Delay Interferometry (TDI)
  (arxiv.org/abs/2105.02054) \cite{TDJ21}.  In so doing we explain
  and quantify our concerns about the results derived by Vallisneri,
  Bayle, Babak and Petiteau \cite{Vallisneri2020} by
  applying their data processing technique (named TDI-$\infty$) to the
  two heterodyne measurements made by a two-arm space-based GW
  interferometer.

  First we show that the solutions identified by the TDI-$\infty$
  algorithm derived by Vallisneri, Bayle, Babak and Petiteau
  \cite{Vallisneri2020} {\underbar {do}} depend on the
  boundary-conditions selected for the two-way Doppler data. We prove
  this by adopting the (non-physical) boundary conditions used by
  Vallisneri {\it et al.} and deriving the corresponding analytic
  expression for a laser-noise-canceling combination. We show it to be
  characterized by a number of Doppler measurement terms that grows
  with the observation time and works for any time-dependent time
  delays. We then prove that, for a constant-arm-length interferometer
  whose two-way light times are equal to twice and three-times the
  sampling time, the solutions identified by TDI-$\infty$ are linear
  combinations of the TDI variable $X$.
  
  In the second part of this document we address the concern expressed
  by Bayle {\it et al.} regarding our matrix formulation of TDI when
  the two-way light-times are constant but not equal to integer
  multiples of the sampling time. We mathematically prove the
  homomorphism between the delay operators and their matrix
  representation \cite{TDJ21} holds in general. By sequentially
  applying two order-$m$ Fractional-Delay (FD) Lagrange filters of
  delays $l_1$, $l_2$ we find its result to be equal to applying an
  order-$m$ FD Lagrange filter of delay $l_1 + l_2$. On physical
  grounds this reflects the fact that sequentially applying two
  order-$m$ interpolators can't result in a higher order ($3m -2$)
  interpolator! We further show that the homomorphism holds in the
  general case of time-dependent arm-lengths for (i) the continuum
  limit and (ii) for fractional delays. In this case the operators do
  not commute. We further argue that the homomorphism extends to the
  general case of time-dependent arm-lengths.
\end{abstract}
\maketitle

\section{Introduction}
\label{SecI}

Time-Delay Interferometry is the data processing technique for
canceling the laser noise from the heterodyne measurements made by
future space-based, unequal-arm gravitational wave (GW)
interferometers. In its simplest-form implementation it entails
properly time-shifting and linearly combining the two two-way Doppler
data measured by an unequal-armlength interferometer so as to cancel
the laser phase fluctuations at any time $t$. TDI is a "local"
operation in that, for a constant-armlength interferometer for
instance, it requires to properly combine four samples of the two
Doppler data selected at times $t$, $t - l_1$ and $t - l_2$, with
($l_1, l_2$) being the times spent by the light to complete a
round-trip within arm \# 1 and \# 2 respectively. In the case of
delay-times changing linearly with time over a time-scale equal to the
round-trip-light-time (RTLT) itself \footnote{This is a dynamic configuration
  resulting from carefully selecting the spacecraft trajectories. It
  has been adopted by LISA and other currently proposed space-based GW
  interferometers to minimize the magnitude of the Doppler beat-notes
  between the received and receiving laser beams so that they fall
  within the photo receiver's operational band-width} the so called
``second-generation'' TDI combination (TDI2) was then derived to account
for the time evolution of the light-times and suppress the laser noise
way below the secondary noises. In this case the laser noise
suppression is achieved by combining eight samples of the two Doppler
data.  As pointed out in \cite{TD2020}, TDI can be extended to any
time-dependent time-delays by applying an iterative procedure in which
the laser noise is effectively suppressed (but not exactly canceled)
to a level many orders of magnitude lower than that defined by the
remaining noise sources affecting the Doppler measurements.

In a recent publication by Vallisneri {\it et al.}
\cite{Vallisneri2020} a new data processing algorithm for canceling
the laser noise from the two two-way Doppler data measured by a two
unequal-length arms space-based detector was proposed. In it the two
sampled Doppler measurements are simultaneously processed by
constructing an array containing the interleaved two Doppler data
measurements. By then applying SVD decomposition to the rectangular
matrix relating this array to the array associated with the laser
noise, they can identify a number $n$ of combinations that are
laser-noise-free (with $n$ being the number of data samples from each
Doppler data). In our article \cite{TDJ21} we pointed
out that the "boundary conditions" (i.e. the relationship between the
two Doppler measurements and the laser noise during the first
RTLTs) adopted in their article were not physically
correct. From there we argued that the solutions found by the
TDI-$\infty$ algorithm necessarily had to reflect their implemented
boundary conditions. Note we did not claim TDI-$\infty$ to be
mathematically incorrect; we rather stated that the solutions
discussed in their article reflected the boundary conditions assumed
and therefore needed to be reanalyzed.

This document is organized as follows. In section \ref{SecII} we
derive a data combination corresponding to the non-physical boundary
conditions. Such a combination is characterized by a number of data
samples that grows with time and works for any time-dependent
light-times. It is easy, however, to show that such a combination
would not cancel the laser noise if applied to the actual
measurements. To understand what the TDI-$\infty$ solutions would look
like when the correct boundary conditions are implemented, we
performed an analytic SVD decomposition (by using the program {\it
  Mathematica}) of the rectangular matrix relating the Doppler to the
laser noise arrays. By assuming the two light-times to be constant and
equal to integer multiples of the sampling time we found the resulting
solutions to be all linear combinations of the TDI measurement $X$.

In section \ref{SecIII} we then turn to our matrix representation of
TDI and mathematically prove that the homomorphism between the space
of the delay operators and their corresponding matrices holds in
general.

\section{The boundary conditions}
\label{SecII}

TDI-$\infty$ establishes a linear relationship between the sampled
Doppler measurements and the laser noise arrays. To understand its
formulation let us consider again the simplified (and stationary)
two-arm optical configuration. In it the laser noise, $C(t)$, folds
into the two two-way Doppler data, $y_1(t)$, $y_2(t)$, in the
following way (where the contributions from all other physical effects
affecting the two-way Doppler data have been disregarded)
\begin{eqnarray}
  y_1(t) =  C(t - l_1(t)) - C(t)  \ ,
\nonumber
  \\
  y_2(t) = C(t - l_2(t)) - C(t)  \ ,
\label{eq1}
\end{eqnarray}
where $l_1$, $l_2$ are the two RTLTs, in general also functions of
time $t$.  Operationally, Eq. (\ref{eq1}) says that each sample of the
two-way Doppler data at time $t$ contains the difference between the
laser noise $C$ generated at a RTLT earlier,
$t - l_i(t) \ , \ i = 1, 2$ and that generated at time $t$.  The
important point to note here is what happens during the first $l_i$
seconds from the instant $t = 0$ when the laser is switched on. Since
the $y_i$ measurements are the result of interfering the returned beam
with the outgoing one, during the first $l_i$ seconds (i.e. from the
moment the laser has been turned on) the $y_i$ measurements are
identically equal to zero because no interference measurements can be
performed during this time. In \cite{Vallisneri2020}, however, only
the first terms on the right-hand-sides of Eq. (\ref{eq1}) were
disregarded during these time intervals.  Although this error has been
minimized by Bayle {\it et al.} in their document, it is rather
relevant to the matter discussed here. And it is even more relevant
for correctly simulating the two-way Doppler data when assessing the
performance of future GW interferometers. Even if we would consider
start processing the Doppler data at any time $t$ after the first RTLT
has past, we would still be confronted by the fact that the Doppler
measurement $y_i$ at time $t$ contains laser noise generated at time
$t$ and at time $t - l_i$. In other words, there exists a
time-misalignment between the array of the Doppler measurement and
that of the laser noise and physical boundary conditions have to be
accounted for.

Although the TDI-$\infty$ technique is mathematically correct, by
implementing with it the physical boundary conditions results in
solutions that are quite different from those associated with the
non-physical boundary conditions. By using the correct boundary
conditions (in the stationary configuration with the RTLTs equal to
twice and three times the sampling time) we have verified analytically
that the $n-6$ observables mentioned by Bayle {\it et al.} are
actually equal to linear combinations of the TDI observable $X$
defined at each of the sampled times. By implementing the non-physical
boundary conditions, on the other hand, one finds solutions equal to
linear combinations of the TDI observable $X$ defined at each of the
sampled times, plus \textit{some additional term} that would not
cancel the laser noise in the measured data. This additional term is a
function of $y_1$ and $y_2$ defined at times $t < l_1, l_2$ and thus,
is a manifestation of the non-physical boundary conditions.

One such a combination can be derived in the following way. Let us
consider the two-way Doppler measurement $y_i(t)$ during the first
RTLT $l_i$, i.e. $t_0 \leq t < t_0 + l_i$ ($t_0$ being the time when
the laser is switched on). Its expression, as  given in
\cite{Vallisneri2020}, is equal to
\begin{equation}
  y_i(t) = -C(t) \ \ \ , \ \ \ t_0 \leq t < t_0 + l_i \ .
  \label{eq2}
\end{equation}
During the second RTLT we then have
\begin{equation}
  y_i(t) = C(t - l_i) - C(t) \ \ \ , \ \ \ t_0 + l_i \leq t < t_0 +
  2l_i \ ,
  \label{eq3}
\end{equation}
which can then be rewritten in the following form after including the  expression for the laser noise
given by Eq. (\ref{eq2})
\begin{equation}
  y_i(t) + y_i(t - l_i) = -C(t) \ \ \ , \ \ \ t_0 \leq t < t_0 +
  2l_i \ .
  \label{eq4}
\end{equation}
It is then easy to derive the following expression for the laser noise
at an arbitrary time $t < t_0 + N l_i$
\begin{equation}
  \rho_i \equiv \sum_{k=0}^{N} y_i(t - k l_i) = -C(t) \ .
\label{C}
\end{equation}
If we now take
the difference $\rho_1 - \rho_2$ we end up canceling the laser noise
at any time $t$ and achieve sensitivity to GWs. Note that the above
derivation can easily be extended to the case in which the RTLTs are
arbitrary functions of time since each Doppler data is delayed by its
own RTLT. Also note the number of Doppler measurements
entering in the combination $\rho_1 - \rho_2$ increases with $t$.
As a final observation, Eq.(\ref{C}) would allow us to measure the noise of a
laser at any time $t$ and therefore have a noiseless laser! 
\vskip 6pt
\noindent
In summary, our findings lead us to the conclusion that the results
quoted in \cite{Vallisneri2020} depend on the boundary conditions
selected by the authors. This is not to say that TDI-$\infty$ is
mathematically incorrect. Rather, it makes it difficult to
understand the nature of the solutions it can find in the general case
of unequal and arbitrarily time-changing arms.  It was the difficulty we experienced to understand what was happening ``under the hood'' of TDI-$\infty$ that
resulted in our matrix formulation of TDI. As stated by Vallisneri {\it et al.} \cite{Vallisneri2020} and by us \cite{TDJ21} a matrix formulation of the technique for canceling the laser noise may provide
computational advantages to the data analysis tasks of space-based GW missions.

\section{The homomorphism: delay operators in matrix avatar}
\label{SecIII}

The homomorphism concept is fundamental and should hold in every
situation of time delays; whether they are integer multiples of the
sampling interval, or fractional or time dependent. Here we argue that
this is indeed so. We have already shown that this is valid for the
case of integer multiples of sampling interval \cite{TDJ21}. As a
matter of principle, one may argue that if the Doppler data could be
sampled at a rate as high as required by TDI (corresponding to a
sampling time of about (10 m/c) sec), then the issue raised on constant fractional delays would not exist and the equality $\phi(D_1 D_2) = \phi(D_1) \phi(D_2)$
would hold. In fact in the continuum limit of the sampling interval
$\Delta t \longrightarrow 0$, the matrix representation of a delay
operator $D_1$ with delay $l_1 (t)$ tends to a delta function
$\delta [t' - (t - l_1 (t))] \equiv \D_1 (t, t')$. Here the matrix $\D_1 (t, t')$ acts on the continuous data stream $y (t)$ as follows:
\be
D_1 y (t) = \int dt'~ \D_1 (t, t') y (t') = \int dt'~ \delta [t' - (t - l_1 (t))]~ y (t') ~=~ y (t - l_1 (t)) \,,
\ee
which is consistent with the usual definition. Here the homomorphism $\phi$ is $\phi (D_1) = \D_1 (t, t')$. If one takes two such
operators even with time-dependent delays $l_1 (t)$ and $l_2 (t)$, and
applies the two operators successively then the result is again a
delta function with a delay $l_1 (t) + l_2 (t - l_1 (t))$ as shown below: 
\bea
\phi (D_1) \star \phi (D_2) &=& (\D_1 \star \D_2)~(t, \tpp) \,, \no \\
&=& \int dt'~ \D_1 (t,t') ~\D_2 (t', \tpp) \,, \no \\
&=& \int dt'~\delta [t' - (t - l_1 (t))]~\delta [\tpp - (t' - l_2 (t'))] \,, \\
&=& \delta [\tpp - \{t - l_1 (t) - l_2 (t - l_1 (t)) \}] \equiv \phi (D_1 D_2) \,.  
\eea
This proves that the matrix representation in the continuum case is
also a homomorphism although,
$$\phi (D_1) \star \phi (D_2) = \D_1 \star \D_2 \neq \D_2 \star \D_1 = \phi (D_2) \star \phi (D_1).$$ 
The operators do not commute in general when the arm lengths are time
dependent. The operators then form a non-commutative polynomial ring,
sometimes also called a free algebra (more about this later). When the
delays are constants, the operators $\D_1$ and $\D_2$ commute and the
operators form a commutative polynomial ring. Thus we have shown that
the homomorphism holds in the continuum limit in addition to the case
of delays being integer multiples of the sampling interval (constant
time-delays) - the opposite end, so to speak.
\par

In practice however, one has nonzero sampling intervals $\Delta t > 0$ (we assume uniform sampling)
and also to avoid oversampling the data one could apply an appropriate
fractional delay filter to the Doppler measurement and achieve
digitally the oversampling mentioned above. To compute fractional
delays if one uses the Lagrange interpolation say on $m$ points, then
one can envisage a $m \times m$ matrix of Lagrange polynomials
$\D(\alpha)$, where $\alpha$ is the delay, acting on the data $y$ (we
write the delays as $\a, \b, ...$ in order to not confuse with the
Lagrange polynomials which are also denoted by $l_i$). If one
considers another delay $\beta$ we have the matrix $\D (\beta)$. If
one uses the {\it same} sample points also for all delays including
the total delay, then one can show that
$\D (\alpha + \beta) = \D (\alpha) \D (\beta)$. This follows from the
properties of Lagrange polynomials. We give below the proof:

For concreteness sake, consider just $m = 3$ points at $t = 0, 1, 2$
and let $l_j (t), ~~j = 0, 1, 2$ be the Lagrange polynomials. We do
not need them explicitly. Let $p (t)$ be the interpolating polynomial
which is required to pass through the points $y_0, y_1, y_2$ at
$t = 0, 1, 2$ respectively. Then we have, \be p (t) = l_0 (t) y_0 +
l_1 (t) y_1 + l_2 (t) y_2 \,.  \ee We just need to use the property of
Lagrange polynomials: \be l_j (t = k) = \delta_{jk} \ee From this we
have $p (k) = y_k$ and so: \be p (t) = l_0 (t) p(0) + l_1 (t) p (1) +
l_2 (t) p (2) \,.  \ee Consider the first term of the product matrix,
$l_0 (\a + \b) \equiv p (\a)$ where $\b$ is held constant. Then at each value of $\a = 0, 1 , 2$ we have $p (k) = l_0 (\b + k)$. Thus we
get: \be l_0 (\a + \b) = \sum_{k = 0}^2 l_k (\a) l_0 (\b + k) \,.  \ee
In general we have: \be l_n (\a + \b) = \sum_{k} l_k (\a) l_n (\b + k)
\,.
\label{eq:add}
\ee
This is in fact the addition theorem for Lagrange polynomials for integer valued nodes, say at $k = 0, 1, ..., m$. The matrix $\D (\a)$ for $m = 2$ is:
\be
\D (\a) =  \left \| \begin{array}{ccc} 
l_0 (\a) & l_1 (\a) & l_2 (\a) \\
l_0 (\a + 1) & l_1 (\a + 1) & l_2 (\a + 1) \\
l_0 (\a + 2) & l_1 (\a + 2) & l_2 (\a + 2) 
\end{array} \right \| \equiv \D_{jk} (\a) \,, 
\ee
where $\D_{jk} (\a) = l_k (\a + j)$. Taking two such matrices corresponding to $\a$ and $\b$ and multiplying them together, we have,
\be
\sum_{k} \D_{jk} (\a) \D_{kn} (\b) = \sum_k l_k (\a + j) l_n (\b + k) \equiv l_n (\a + \b + j) = \D_{jn} (\a + \b) \,,
\label{hom}
\ee where we have used the addition theorem in
Eq. (\ref{eq:add}). Although we just used 3 time stamps the results
are true for $m$ points. Also one might think, that since the product
of Lagrange polynomials appears as entries in the product of the
matrices, it might lead to polynomials of degree $2m - 2$ (we do not understand $3m -2$). But this
does not happen, as the addition theorem shows; the terms of degree
greater than $m - 1$ cancel out, leaving behind a $m - 1$ degree
polynomial.

In practice, choosing the same set of sample points may not be
feasible for delays much greater than the sampling interval and so
different sets of sample points must be chosen for different delays
but then the matrices may {\it appear} different. But then care must
be taken to translate the matrices to a common reference in order to
compare them. Then the closure property of the polynomials can be
explicitly seen to hold. Since here we are concerned about matters of principle, we may choose $m$ sufficiently large to cover all delays.

We would like to emphasize that Eq. (\ref{hom}) is valid for time
dependent delays also. Both $\a$ and $\b$ become now functions of
time. If one applies the delay $\b$ first and then $\a$, the
combined delay is $\a + \b (\a) \equiv \a \oplus \b$ and in the
reverse case it is $\b + \a (\b) = \b \oplus \a$ which are in general
unequal. Then we have the situation: \be 
\D (\a \oplus \b) = \D [\a +
\b (\a)] = \D (\a) \D [\b (\a)] \neq \D (\b) \D [\a (\b)] = \D [\b +
\a (\b)] = \D (\b \oplus \a) \,.
\label{hom_timedep}
\ee 
Eq. (\ref{hom_timedep}) shows that the homomorphism
also holds for time-dependent fractional delays with non-commuting
operators.
\par

It is possible that some other interpolation method may be employed
for obtaining the fractional delay filter. In that case, the
fractional delay filter may only produce an approximation to the true
matrix representation of the operator albeit very accurate. This
approximation may result in producing $\D(\a) \D (\b)$ different from
$\D (\a + \b)$ making it {\em appear} that it is not a
homomorphism. But both $\D(\a) \D (\b)$ and $\D (\a + \b)$ are two
{\it different} approximations to the same exact matrix, which exists
in principle and also perhaps realized in practice with a higher
sampling rate. Therefore, while applying approximation methods, their
limitations must be kept in mind.

\section{Time dependent arm lengths and non-commutative polynomial rings}

Although in reference \cite{TDJ21} we have mostly dealt with constant
arm length delays, in section IV A we have described the general ring
structure of the operators with time dependent arm-lengths. We then
have a non-commutative ring of operators. When discussing
homomorphisms, our presentation in \cite{TDJ21} might have been
misconstrued that we were only referring to the situation of 3
constant arm-lengths which was first addressed by Dhurandhar, Nayak
and Vinet (2002) \cite{DNV02}. As correctly pointed out by Bayle {\it
  et al.} the TDI observables defined in \cite{DNV02} cannot deal with
time-dependent arm-lengths, because the operators commute. But we
would like to point out that since then more work has been
done. Ref.\cite{NV04} obtains the first module of syzygies for 6
constant arm-lengths with 6 commuting operators, where one has a
rigidly rotating LISA triangle and the Sagnac effect makes the up and
down links between spacecraft unequal. Here too the ring is
commutative. This is the so-called TDI 1.5. The problem of TDI with
time-dependent arm-lengths has been addressed in \cite{SVDsymp} and
\cite{DNV10}. In this case we have a non-commutative polynomial ring
which again leads to the first module of syzygies. Here the operators
do not commute and any element of the ring is a word or a string of
operators, and the order of the operators is important. For example,
if $x$ and $y$ are the operators, then there are 4 distinct monomials
$x^2, xy, yx, y^2$ of second degree, instead of 3 in the commutative
case. At third degree there are 8 distinct monomials and so on. In
this general case, obtaining the TDI variables is a very difficult
problem. It is not even clear whether the Gr\"obner basis is
finite. Nevertheless, partial approximate solutions for the special
case of only two arms of LISA functioning and with arm-lengths slowly
varying, have been obtained. The solutions are of the Michelson type
\cite{DNV10} and are TDI-2. All the above literature has been reviewed
comprehensively in \cite{TD2020}.

In summary, the matrix representation of the delay operators is
correct and it captures all the well known properties of the TDI space
and its generators.

\bibliographystyle{apsrev}
\bibliography{refs}
\end{document}